# On the Origin of the Double Superconducting Transition in Overdoped $YBa_2Cu_3O_x$

R. Lortz[1], T. Tomita[2], Y. Wang[1], A. Junod[1], J.S. Schilling[2], T. Masui[3], S. Tajima[3]

[1] Department of Condensed Matter Physics, University of Geneva,
24 Quai Ernest-Ansermet, CH-1211 Geneva 4, Switzerland

[2] Department of Physics, Washington University
CB 1105, One Brookings Dr., St. Louis, MO 63130, USA

[3] Department of Physics, Osaka University,
Toyonaka, Osaka 560-0043, Japan.

The superconducting transition in a single overdoped, detwinned $YBa_2Cu_3O_x$ (YBCO) crystal is studied using four different probes. Whereas the AC and DC magnetic susceptibilities find a dominant transition at 88 K with a smaller effect near 92 K, the specific heat and electrical resistivity reveal only a single transition at 88 K and 92 K, respectively. Under hydrostatic pressures to 0.60 GPa these two transitions shift in opposite directions, their separation increasing. The present experiments clearly show that the bulk transition lies at 88 K and originates from fully oxygenated YBCO; the 92 K transition likely arises from filamentary superconductivity in a minority optimally doped phase (< 1 %) of YBCO located at or near the crystal surface.

**Keywords:** high pressure, transition temperature variations, $YBa_2Cu_3O_x$, high-temperature superconductivity, surface superconductivity, specific heat, thermodynamic properties, transport properties

**PACS numbers: 74.62.-c Transition temperature variations, 74.25.Bt Thermodynamic properties, 74.25.Fy Transport properties, 74.62.Fj Pressure effects**

# Introduction

A double superconducting transition has frequently been reported in overdoped YBCO single crystals [1,2] and ceramics [3,4,5,6,7]. Some studies have been used to support the contention that the double transition is an intrinsic property of a multicomponent superconducting order parameter [5], whereas others attribute the effect to phase-separation in YBCO at different doping levels [1,3,4,6]. It is well known that the dependence of $T_c$ on pressure can give valuable information on the nature of the superconductivity [8,9]. To our knowledge, studies of the effect of high pressures on the double transition have not yet been carried out.

In the present experiments the temperature-dependent specific heat, AC and DC susceptibility, and DC electrical resistivity are measured on a high quality fully oxygenated YBCO single crystal. As is amply illustrated in the present study, these techniques differ widely in their sensitivity to bulk versus filamentary superconductivity. Several examples can be found in the literature where the determination of the superconducting transition temperature $T_c$, or of the upper critical field $H_{c2}$, depends strongly on the method used [10,11]. One prominent example is the persistence of superconductivity in clean samples in the region adjacent to an insulator/metal interface up to a field $H_{c3}(T) \cong 1.7 H_{c2}(T)$ parallel to the surface [12]. A bulk measurement such as the specific heat is insensitive to this effect, as the volume involved in the surface transition is vanishingly small. Transport measurements such as the electrical resistivity may be sensitive to a percolative path through the sample which persists above the bulk $T_c$. The AC or DC magnetic susceptibility is normally a much better measure of the bulk properties, but may exhibit some sensitivity to percolation effects, particularly when they originate from minority phases at or near the sample surface.

In the present studies the value of $T_c$ of the high-temperature superconductor YBa$_2$Cu$_3$O$_7$ (YBCO) as determined by the above four techniques varies by 4-5 K. Some of the methods show a double transition, while others are only sensitive to one of them. Under the application of pressure the upper $T_c$ is enhanced, while the lower $T_c$ decreases. From an analysis of the present results we conclude that the upper $T_c$ is most probably due to percolative paths near the crystal surface from a trace minority phase of optimally oxygen doped YBCO. The lower $T_c$ is clearly identified by the specific heat as characteristic of bulk superconductivity; its pressure dependence $dT_c/dP$ = -1.20(3) K/GPa is shown to be compatible with that anticipated for fully oxygenated YBCO. Macroscopic inhomogeneity can be excluded as origin of the double superconducting transition: small pieces broken from the large piece all show the same double transition. A clear indication of the crystal's quality is furthermore its wide reversible domain in the *H-T* phase diagram with sharp pronounced specific-heat peaks at the first-order vortex-melting transition, as will be reported elsewhere [13].

## Experiment

A high quality, detwinned overdoped YBa$_2$Cu$_3$O$_7$ single crystal was used for the present study. It was oxygenated at 400 °C in 90 bar O$_2$ pressure for 200 hours to obtain essentially full oxygenation of the CuO chains. The superconducting transition temperature $T_c$ at ambient pressure was determined by four different methods: DC electrical resistivity, AC and DC magnetic susceptibility, and DC specific heat which, in this order, are increasingly representative of the bulk volume. The DC resistivity was determined with a standard four-probe technique. A commercial SQUID magnetometer was used to measure the DC susceptibility in a field of 10 G. The specific heat was studied in an adiabatic calorimeter. Specific-heat measurements under pressure have been performed with an AC technique in a quasihydrostatic Bridgman-type pressure cell to 10 GPa [14] on a small piece cut from the main crystal. The AC susceptibility was measured in a He-gas pressure cell on a second small piece at both ambient pressure and under purely hydrostatic pressures to 0.60 GPa. For further details of the He-gas high pressure technique, see Ref. [15].

## Results of Experiment and Discussion

Fig. 1 shows the results of three measurements used to determine $T_c$. In Fig. 1(a) the DC resistivity is plotted versus temperature; the transition midpoint is located at 92 K, with the superconducting onset at 94 K (see right inset). In the left inset in Fig. 1(a) a tiny second anomaly is located near 88 K. In contrast, as seen in Fig. 1(c), the middle of the specific-heat jump is located at 88 K with no anomaly whatsoever visible at 92 K where the dominant resistive transition occurs. The specific heat result shows unequivocally that the bulk transition is at 88 K, a reasonable value for an overdoped YBCO sample. The dominant resistive transition near 92 K must be percolative in nature. These conclusions are supported by the DC and AC susceptibility studies in Fig. 1(b) and Fig. 2, respectively, where the main transition is seen to occur at 88 K (midpoint) with a much smaller transition near 92 K with an onset at 94 K. The DC susceptibility was measured under zero-field cooled (ZFC) and field cooled (FC) conditions. The small applied field of 10 G was chosen sufficiently small to not broaden the superconducting transitions. Interestingly, the onset of irreversibility where the two curves merge is connected with the onset of the upper transition at 94 K.

In order to further characterize the two superconducting transitions we examined their pressure dependences using both the specific heat and AC susceptibility. The measurements of the AC susceptibility are shown in Fig. 2 at ambient and 0.60 GPa hydrostatic pressure; interestingly, it is seen that the pressure derivatives of the two transitions have opposite sign. In Fig. 3 $T_c$ for both transitions is plotted versus pressure. Whereas the filamentary transition at 92 K increases with pressure at the rate +0.98(3) K/GPa, the lower bulk transition decreases at the rate -1.20(3) K/GPa, in good agreement with earlier quasihydrostatic specific heat measurements on the same crystal which are also included in Fig. 3 [14]. In the high pressure specific heat studies, no signature of a second transition above the bulk $T_c$ could be resolved at any pressure [14], confirming the result from the measurement performed under ambient pressure.

Oxygen ordering effects have been shown to have a sizeable influence on $T_c(P)$ at most oxygen doping levels in YBCO [16,17,18,8]. One can check for these effects through experiments where the pressure is applied at room temperature, but released at low temperature (60 K, for example); if $T_c$ does not return to its original zero-pressure value, then oxygen ordering effects are indicated. The mechanism behind this effect is that oxygen anions in the CuO chains have considerable mobility at room temperature and are able to order in response to an applied pressure, thus making a sizeable contribution to a shift in $T_c$. If the pressure is released, however, at low temperatures, the oxygen anions are frozen in place and cannot return back to their original positions, so that $T_c$ does not return to its initial value.

The results of the He-gas measurements on the YBCO crystal are shown in detail in Fig. 3. Following the determination of the $T_c$ values at ambient pressure (point 1), 0.60 GPa hydrostatic pressure was applied at room temperature resulting in shifts in $T_c$ (points 1→2). The pressure was then fully released at 60 K (point 3); both the upper and lower transitions are seen to completely revert to their previous values at point 1. This negative result implies that there are no measurable oxygen ordering effects in $T_c$ in the present sample. The final data point (point 4) was obtained after applying 0.31 GPa pressure at 80 K.

The complete absence of oxygen ordering effects in $T_c$ for YBCO is quite unusual. This is believed to occur only under two circumstances: (1) the CuO chains are completely filled with oxygen so that no ordering is possible (sample has stoichiometry $YBa_2Cu_3O_{7.0}$), or (2) the sample is optimally doped (sample has stoichiometry $YBa_2Cu_3O_{6.95}$), placing $T_c$ at its maximum value ~ 92 K, an extremum, so that small changes in carrier concentration due to oxygen ordering cause no change in $T_c$. Note that no oxygen ordering would be possible in a sample where the CuO chains are completely devoid of oxygen anions, i.e. the sample has stoichiometry $YBa_2Cu_3O_{6.0}$; however, such a sample is not superconducting, but rather an insulator.

Based on the above results, a possible scenario for the double superconducting transition in the present sample is that the bulk transition at 88 K originates from a completely oxygenated (overdoped) YBCO crystal, whereas the filamentary transition at 92 K is caused by a trace minority phase of optimally doped YBCO. In fact, the latter value of $T_c$ and its pressure derivative, $dT_c/dP$ = +0.98(3) K/GPa, agrees well with the results of numerous high-pressure studies of *bulk* superconductivity on optimally doped YBCO samples [16,19,18,8]. A negative derivative $dT_c/dP$ is also expected, and has been observed [16,19], for overdoped YBCO samples.

A further test of the above scenario is provided by ascertaining whether the magnitude of the pressure derivative for the bulk transition at 88 K, namely $dT_c/dP$ = -1.20(3) K/GPa, is consistent with this transition being at a temperature 4 K below the value $T_c$ = 92 K appropriate for optimal doping. To explore this question, consider the following canonical "inverted parabolic" dependence of $T_c(n)$ for YBCO and other high-temperature superconductors on the hole carrier concentration $n$ per Cu cation in a single $CuO_2$ sheet

$$T_c = T_c^{max}\left[1 - A(n - n_{opt})^2\right] \qquad (1)$$

where A = 82.6 and $n_{opt}$ = 0.16 [20,21]. At optimal doping, $n = n_{opt}$ and $T_c = T_c^{max}$ = 92 K. Now if the doping level $n$ is increased above optimal doping by increasing the oxygen concentration until $T_c$ = 88 K, then it follows from Eq. (1) that $(n-n_{opt})$ = 0.0229. Taking the derivative of Eq. (1) with pressure, assuming that $A$ and $n_{opt}$ are not pressure dependent, we obtain

$$\frac{dT_c}{dP} = \frac{dT_c^{max}}{dP}\left[1 - A(n - n_{opt})^2\right] - 2AT_c^{max}(n - n_{opt})\frac{dn}{dP}. \qquad (2)$$

From studies on YBCO at variable oxygen doping level it has been demonstrated that $dT_c^{max}/dP \approx$ +1 K/GPa [16,19,18,8]. After substituting in Eq. (2) the above values of $A$, $T_c^{max}$, $dT_c^{max}/dP$, $(n - n_{opt})$, and $dT_c/dP$ = -1.20 K/GPa, we solve for the pressure derivative of the hole carrier concentration to obtain $dn/dP$ = +0.0061 holes/GPa. This value is in excellent agreement with the results of previous studies where $T_c$ and its pressure derivative were determined for YBCO over a wide range of doping level [19,22]. This lends further support that the above scenario is a valid description of the double superconducting transition.

The present specific heat and magnetic susceptibility studies clearly demonstrate that the bulk superconducting transition in the present overdoped YBCO crystal occurs at 88 K, the transition near 92 K being filamentary. The above analysis demonstrates that these values of $T_c$ and their pressure derivatives are consistent with the minority filamentary phase with $T_c$ = 92 K having a slightly lower oxygen content than in the bulk. The magnitude of the filamentary transition in the magnetic susceptibility at 92 K is seen in Figs. 1(b) and Fig. 2 to be 10-15% of the total transition. Were the actual volume fraction of the filamentary phase to be this high, an anomaly would definitely have been seen in the specific-heat measurement. The fact that no anomaly whatsoever is seen in the specific heat at 92 K indicates that the volume fraction of the filamentary phase must be less than approximately 1%. Such a large shielding effect as 10-15% in the magnetic susceptibility would be most readily possible if the filamentary phase were concentrated in the outer regions (surface) of the YBCO crystal. This conclusion is further supported by the fact that the onset of irreversibility as determined by the FC and ZFC experiment in the DC susceptibility in Fig. 1b is associated with the upper filamentary transition. Furthermore a peak in $\chi^{//}$ (see Fig. 2) indicates that macroscopic shielding currents already start to circulate at the upper transition temperature. A slightly lower oxygen content at the surface than in the bulk would occur if, following full oxygenation of the bulk sample at high temperatures and high oxygen pressures, some oxygen at and near the surface slowly diffuses out, leaving behind a slightly underoxygenated surface layer with the higher value of $T_c$ = 92 K, corresponding to optimal doping. For the specific-heat measurements under pressure [14] a piece of the main sample was polished on all sides within a few hours before closing the pressure cell. The original surface layer was thus completely removed. In this case the resistive transition occurred at the same temperature as the transition in the specific heat, confirming the above scenario.

The physics of the present double superconducting transition is in good agreement with that observed previously in single crystals [1], but differs somewhat from that in polycrystalline samples. In the latter case the onset of irreversibility was reported to be connected with the lower transition. A higher oxygenation of the grain surface was made responsible, preventing further oxygenation of the interior of the grains during the oxygenation process [3]. Due to the grain structure of the polycrystalline samples, the surface has a much larger contribution to the volume, thus explaining the sensitivity of bulk measurements to both transitions. In single crystals the reverse is true and the surface contribution can be neglected compared to the bulk transition. However, the present experiments show that care must be taken when $T_c$ is determined by surface sensitive experiments in overdoped single crystals.

**Acknowledgments.** This work was partly supported by the Swiss National Science Foundation through the National Center of Competence in Research "Materials with Novel Electronic Properties-MaNEP". The research at Washington University is supported by NSF grant DMR-0404505. The authors thank T. Plackowski and E. Giannini for help in the sample annealing and characterization.

# Figure Captions

**Figure 1.** Three different methods used to determine $T_c$ in overdoped YBCO crystal. Solid lines are guides to the eye. a) DC electrical resistivity along *a*-axis showing main superconducting transition at 92 K with onset at 94 K (right inset) and a tiny feature at 88 K (left inset); b) DC magnetic susceptibility at 10 G for both zero-field cooling (ZFC) and field cooling (FC) with main transition at 88 K (midpoint) and small transition at 92 K (onset at 94 K, as seen in inset); c) specific heat $C/T$ measured in an adiabatic calorimeter using the entire 53 mg crystal showing bulk transition at 88 K, but no visible anomaly at 92 K (here 1 g at = 1/13 mole).

**Figure 2.** Real (upper) and imaginary (lower) parts of the AC susceptibility (1023 Hz, $H_{rms}$ = 0.1 G) versus temperature at ambient and 0.60 GPa hydrostatic pressure for a small piece from same YBCO crystal as used for the measurements in Fig. 1. Data in lower graph are enhanced four times and shifted vertically for clarity.

**Figure 3.** $T_c$ versus pressure as derived from AC susceptibility ($\chi'$) data (see Fig. 2) and specific heat data from Ref. [14]. Numbers give order of measurements. Solid lines are guides to eye.

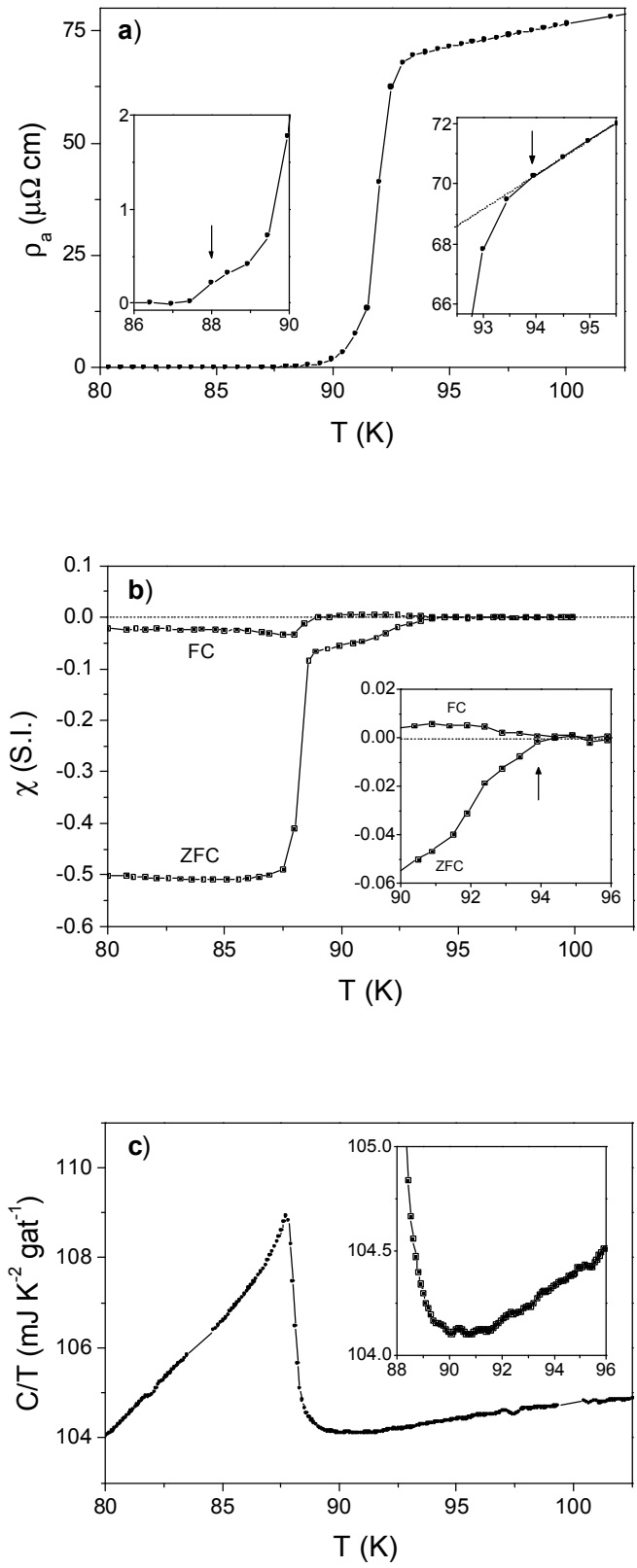

Figure 1

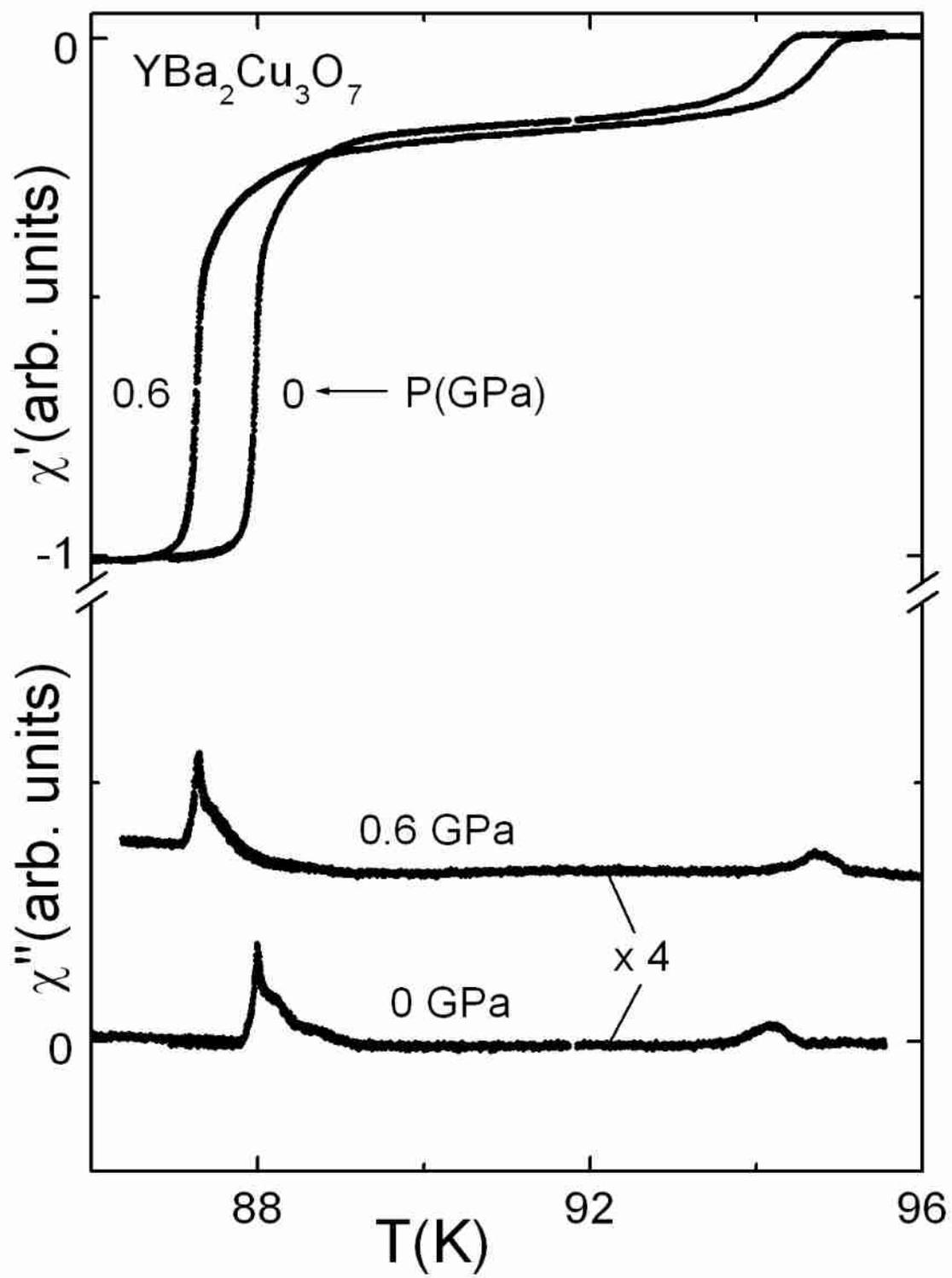

**Figure 2**

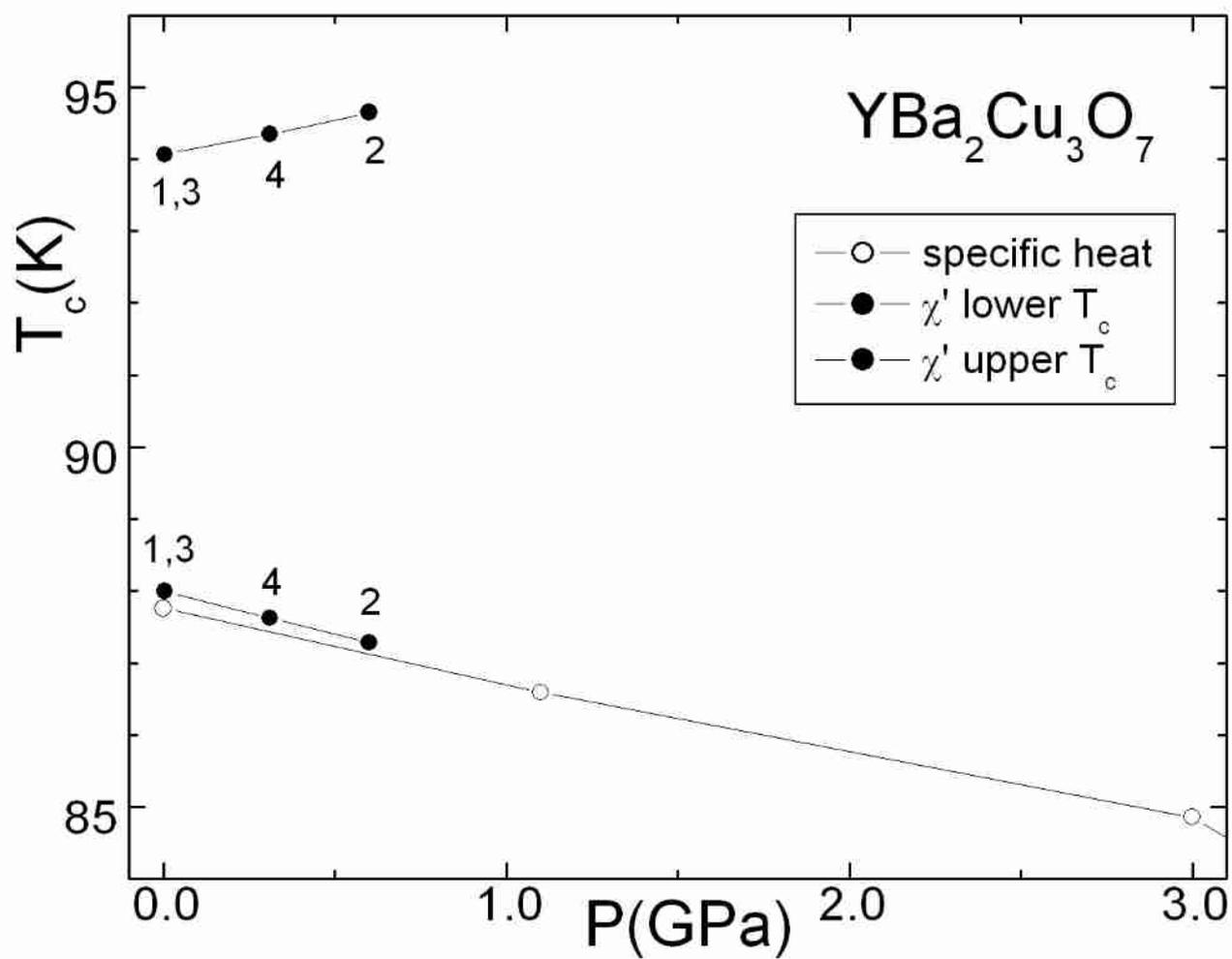

**Figure 3**